\documentclass[conference]{IEEEtran}
\IEEEoverridecommandlockouts
\usepackage{cite}
\usepackage{amsmath,amssymb,amsfonts}
\usepackage{algorithmic}
\usepackage{graphicx}
\usepackage{textcomp}
\usepackage{xcolor}
\usepackage{tabularx}
\usepackage[binary-units]{siunitx}
\usepackage{xcolor,colortbl}
\usepackage{hyperref}
\usepackage{fancyhdr}
\usepackage{lipsum}
\usepackage[T1]{fontenc}
\usepackage[utf8]{inputenc}
\usepackage{authblk}
\usepackage{lipsum}
\usepackage[american]{babel}

\usepackage[nohyperlinks,nolist]{acronym}
\acrodef{AI}[AI]{artificial intelligence}
\acrodef{DL}[DL]{deep learning}
\acrodef{ANN}[ANN]{artificial neural network}
\acrodef{RNN}[RNN]{recurrent neural network}
\acrodef{DNN}[DNN]{deep neural network}
\acrodef{SNN}[SNN]{spiking neural network}
\acrodef{CNN}[CNN]{convolutional neural network}
\acrodef{GPU}[GPU]{graphics processing unit}
\acrodef{BPTT}[BPTT]{back-propagation through time}
\acrodef{FPTT}[FPTT]{Forward Propagation Through Time}
\acrodef{ULP}[ULP]{ultra-low power}
\acrodef{CIM}[CIM]{compute-in-memory}

\DeclareSIUnit\op{Op}
\DeclareSIUnit\inf{inf}
\DeclareSIUnit\sy{Sy}
\def\BibTeX{{\rm B\kern-.05em{\sc i\kern-.025em b}\kern-.08em
    T\kern-.1667em\lower.7ex\hbox{E}\kern-.125emX}}
\begin{document}
\bstctlcite{BSTcontrol}

\author[1]{Manil Dev Gomony}
\author[1]{Floran de Putter}
\author[4]{Anteneh Gebregiorgis}
\author[3]{Gianna Paulin}
\author[2]{Linyan Mei}
\author[2]{Vikram Jain \\}
\author[4]{Said Hamdioui}
\author[1]{Victor Sanchez} 
\author[5]{Tobias Grosser}
\author[1]{Marc Geilen}
\author[2]{Marian Verhelst}
\author[8]{Friedemann Zenke \\}
\author[3]{Frank Gurkaynak}
\author[1]{Barry de Bruin}
\author[1]{Sander Stuijk}
\author[9]{Simon Davidson}
\author[1]{Sayandip De}
\author[10]{Mounir Ghogho \\}
\author[11]{Alexandra Jimborean}
\author[1]{Sherif Eissa}
\author[3]{Luca Benini}
\author[6]{Dimitrios Soudris}
\author[4]{Rajendra Bishnoi \\}
\author[5]{Sam Ainsworth}
\author[1]{Federico Corradi}
\author[10]{Ouassim Karrakchou}
\author[7]{Tim Güneysu}
\author[1]{Henk Corporaal}

\affil[1]{Eindhoven University of Technology}
\affil[2]{Katholieke Universiteit Leuven}
\affil[3]{ETH Zurich}
\affil[4]{Delft University of Technology}
\affil[5]{University of Edinburgh}
\affil[6]{National Technical University of Athens}
\affil[7]{Ruhr-Universität Bochum}
\affil[8]{Friedrich Miescher Institute}
\affil[9]{The University of Manchester}
\affil[10]{International University of Rabat}
\affil[11]{University of Murcia}


\title{CONVOLVE: Smart and seamless design of smart edge processors\\}

\maketitle
\begin{abstract}
With the rise of \ac{DL}, our world braces for \ac{AI} in every edge device, creating an urgent need for edge-AI SoCs. This SoC hardware needs to support high throughput, reliable and secure AI processing at \ac{ULP}, with a very short time to market. With its strong legacy in edge solutions and open processing platforms, the EU is well-positioned to become a leader in this SoC market. However, this requires AI edge processing to become at least 100 times more energy-efficient, while offering sufficient flexibility and scalability to deal with AI as a fast-moving target. Since the design space of these complex SoCs is huge, advanced tooling is needed to make their design tractable. The CONVOLVE  project (currently in Inital stage) addresses these roadblocks. It takes a holistic approach with innovations at all levels of the design hierarchy.  Starting with an overview of  SOTA DL processing support and our project methodology, this paper presents 8 important design choices largely impacting the energy efficiency and flexibility of DL hardware. Finding good solutions is key to making smart-edge computing a reality.
\end{abstract}


\section{Introduction} 

As the world braces for smart applications powered by AI in almost every edge device, there is an urgent need for ultra-low-power (ULP) edge AI System-on-Chips (SoC) or Smart Edge Processor (SEP) that offloads the computing closer to the source of data generation to address the limitations (e.g., latency, bandwidth) of cloud or centralized computing. Based on the current projections, the SEP market is expected to grow about 40\% per year, beyond 70 Billion USD by 2026. 
In contrast to cloud computing, edge AI hardware is far more energy constrained. Hence, low-cost application specific ULP SoCs are needed to make the edge intelligent. The strong ULP requirements can only be achieved by combining innovations from all levels of the design stack, from AI deep learning models, compilers, architecture, micro-architecture, to circuits and devices. Innovations include ULP memristive circuits, exploiting Compute-in-Memory (CIM) and approximate computing, more advanced DL models, online learning, exploiting dynamism and reconfiguration at DL-, Architecture- and Circuit-levels, while rethinking the whole compiler stack.
 This results in extremely complex edge systems. Therefore, a single framework is needed that ties the innovations from the different levels together to fast design and design-space-exploration (DSE). 
 Hence, we define the main objectives as follows: \emph{(1)~Achieve 100x improvement in energy efficiency compared to state-of-the-art COTS solutions}. \emph{(2)~Reduce design time by 10x to be able to quickly implement an ULP edge AI processor combining innovations from the different levels of stack for a given application}. To understand how we can accomplish the key objectives, we present this paper with the following contributions:
\begin{itemize}
\item Summary of state-of-the-art low power microprocessors for deep learning applications (Sec.II).
\item CONVOLVE three-pillar design methodology that includes design-space exploration and compilation flow that reduces the design time by 10X (Sec.III).
\item Key design choices to be considered for improving energy efficiency by a factor of 100X (Sec.IV-VI).
\end{itemize}


\section{State-of-the-art Edge-AI processing}


Edge-AI applications require high performance and flexible SoCs to efficiently map a diverse set of workloads. Heterogeneous multi-core SoCs can provide such duality by utilizing highly energy-efficient specialized hardware accelerators, possibly supporting different operand precisions. In order to judge the SotA for Edge-AI processing recent SOCs (including several from CONVOLVE partners) and ULP processing approaches are presented.

\begin{figure}[t]
\centering
    \includegraphics[width=0.8\columnwidth]{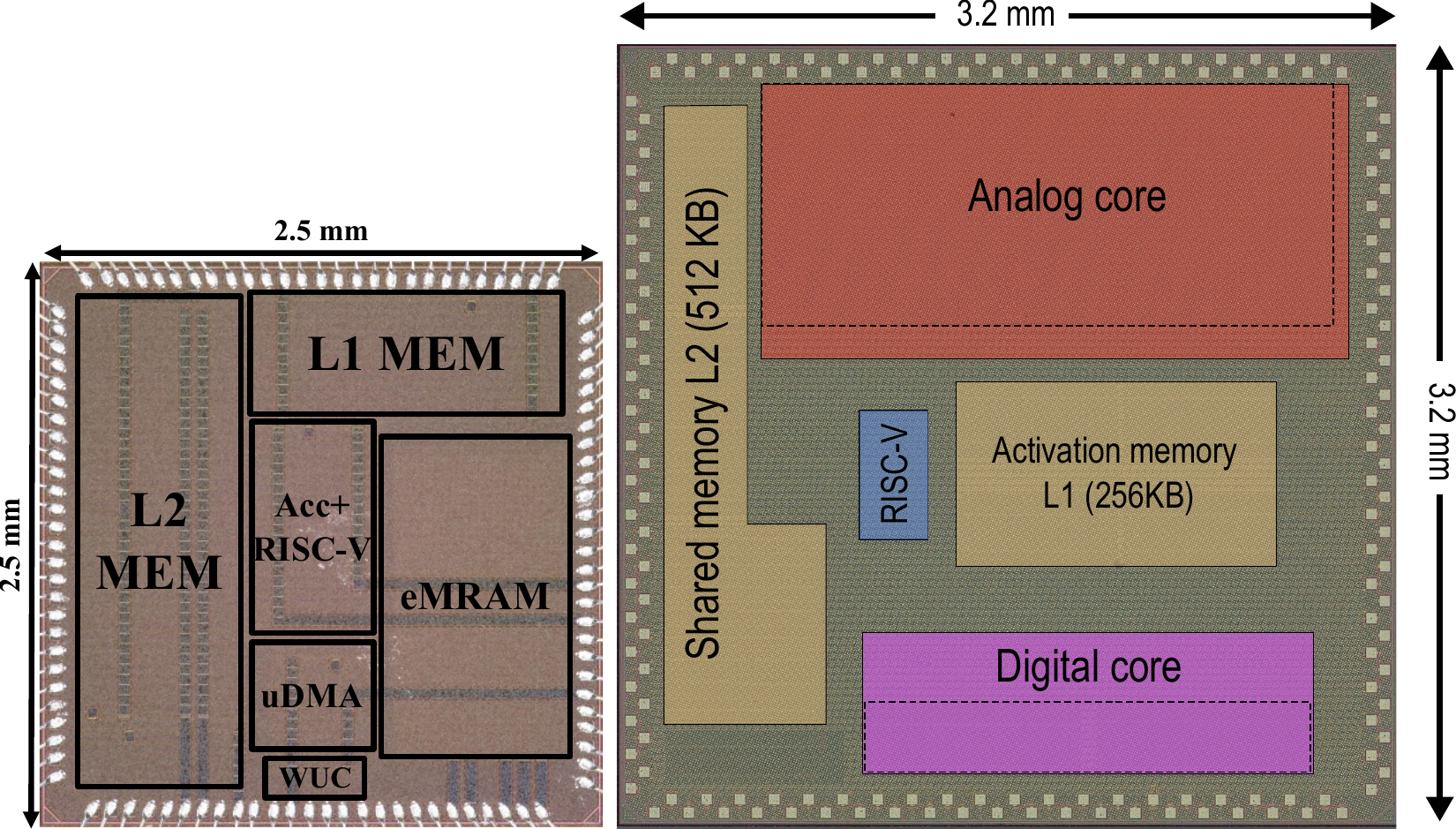}
    \caption{TinyVers (left) and DIANA (right) die micrographs.}
    \label{fig:tv_diana}
\end{figure}

\textbf{\emph{TinyVers} - embedding MRAM}: 
TinyVers~\cite{tinyVers} (Fig.~\ref{fig:tv_diana}) integrates a highly flexible-precision scalable digital accelerator, with a single core RISC-V processor, a power management unit and an eMRAM, to provide a complete standalone edgeAI solution. The accelerator supports diverse AI layer types from DNN (CNN, FC, TCN, GAN, AE) to traditional ML models like SVM at INT2/4/8 precisions. Fabricated in 22nm FDX, it provides 0.8-17 TOPS/W with power consumption ranging from 1.7 $\mu$W in deep sleep to sub-mW when running real AI workloads. 

\textbf{\emph{DIANA} - mixed-signal, mixed-precision}: 
DIANA~\cite{diana} (see Fig.~\ref{fig:tv_diana}) extends the idea of heterogeneity by combining an ULP analog in-memory core with a precision-scalable digital NN accelerator, an optimized shared-memory subsystem and a RISC-V host processor to achieve SotA end-to-end inference at the edge. The SoC achieves peak energy efficiencies of 600 TOPS/W (7bit I, ternary W, 6bit O) for the AIMC core and 14 TOPS/W (8bit I/W/O) for the digital accelerator. When end-to-end ResNet20/CIFAR-10 and ResNet18/ImageNet classification workloads are mapped on the chip, 7 TOPS/W and 5.5 TOPS/W efficiencies are reported at system level respectively.

\textbf{\emph{BrainTTA} - Flexible AI support}: 
A popular method to achieve extremely high energy efficiency is to perform aggressive quantization i.e., reduce operand bit widths to as low as a single bit. However, this may not always be  optimal in terms of accuracy. Thus, a typical edge AI workload consists of various precision levels and layer dimensions.

\begin{figure}[t]
    \centering
    \includegraphics[angle=90,width=0.7\columnwidth]{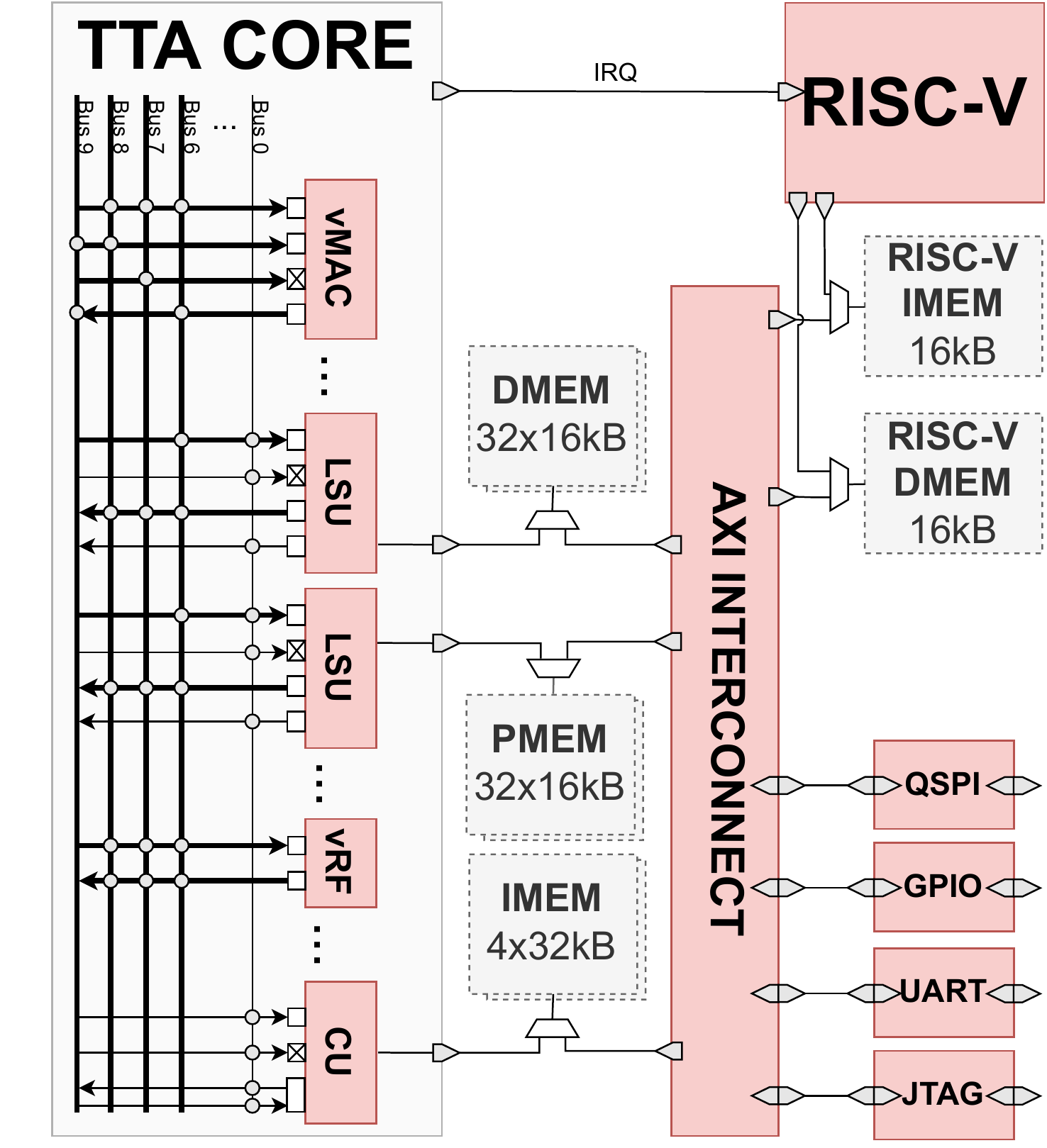}
    \caption{Block diagram of the BrainTTA SoC. The AXI-interconnect forms the border between the RISC-V host and the flexible TTA AI accelerator.}
    \label{fig:braintta}
\end{figure}

BrainTTA [ref] is able to efficiently map various typical AI workloads, because of its inherent flexible datapath from the Transport-Triggered Architecture (TTA). As illustrated in Fig. \ref{fig:braintta}, the SoC consists of a RISC-V processor and a TTA-based accelerator. The accelerator is fully-programmable and is supported by a C-compiler, which greatly simplifies mapping various AI (and other) workloads. BrainTTA, fabricated in 22nm FDX, has a peak energy efficiency of 29/15/2 TOPS/W (binary, ternary, and 8-bit precision) and a throughput of 614/307/77 GOPS.

\begin{figure}[t]
    \centering
    \includegraphics[width=\columnwidth]{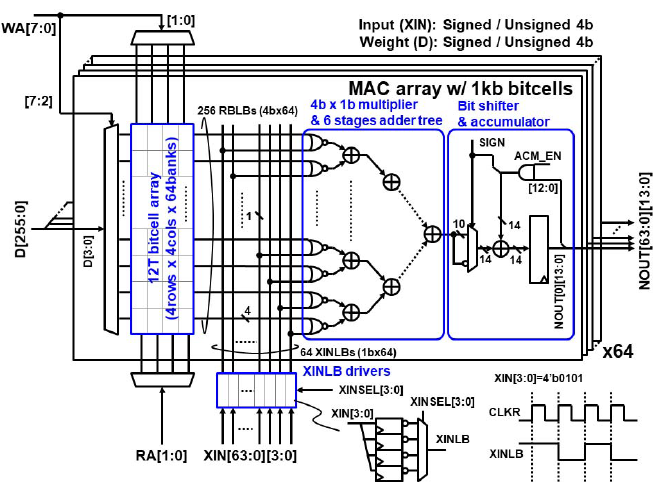}
    \caption{Overall architecture of Digital CIM Macro~\cite{9731754}.}
    \label{fig:digcim}
\end{figure}

\textbf{Digital CIM, SRAM-based}: 
Computing in memory (CIM) has been proposed as a paradigm capable of overcoming the memory-wall problem of traditional computing architectures. The input vector and weight matrix multiplication (i.e. MAC operations), is carried in the analog or digital domain within the memory sub-array. This leads to significant improvements in throughput and energy efficiency. CIM can be realized using standard SRAM as well as emerging non-volatile memory. SRAM-based CIM provides faster write speed and lower write energy with almost unlimited endurance~\cite{si2019twin}. Digital MAC operation in SRAM-based CIM is performed by modifying the memory macro to add the required logic components such as multiplier, shift logic and accumulator unit in the periphery. 
A digital CIM~\cite{9731754}, shown in Fig.~\ref{fig:digcim}, using 12T bitcell supporting wide-range dynamic voltage-frequency scaling (0.5V-0.9V) and flexible precision (4-b and 8-b) MAC operations has an area efficiency of $221~TOPS/mm^2$ (4b), $55~TOPS/mm^2$ (8b), and energy efficiency of $254~TOPS/W$ (4b) and $63~TOPS/W$ (8b).  

\textbf{Analog CIM, RRAM Based} 
 Resistive memories store analog values in the form of resistances, however the surrounding data communication remains digital~\cite{roy2019towards, singh2021srif, singh2021low, new2021unbalanced, bengel2022reliability}. Quantization of analog output to digital data streams is done using an Analog-to-Digital Converter (ADC); it largely determines the overall efficiency of the architectures~\cite{mayahinia2022voltage,diware2022accurate}. An RRAM-based CIM macro with voltage-regulating current sense topology is proposed to improve the area and energy efficiency (~27 TOPS/W) of the ADC design \cite{ spetalnick202240nm}. Moreover, NeuRRAM architecture, which is a 48-core RRAM based CIM hardware, proposes a variable computation bit-precision while performing ADC at low power consumption and compact-area footprint, and achieves the energy efficiency of around 40 TOPS/W \cite{ wan2022compute}. Furthermore, a 195.7 TOPS/W is reported using RRAM-based CIM macro supporting a 8b-input and 8b-weight MAC operations \cite{ xue202116}. This architecture includes an asymmetric group-modulated input scheme to reduce the computing latency as well as a weighted current-to-voltage signal stacking converter for the MAC operations.

\begin{figure}[b]
    \centering
    \includegraphics[width=0.95\columnwidth]{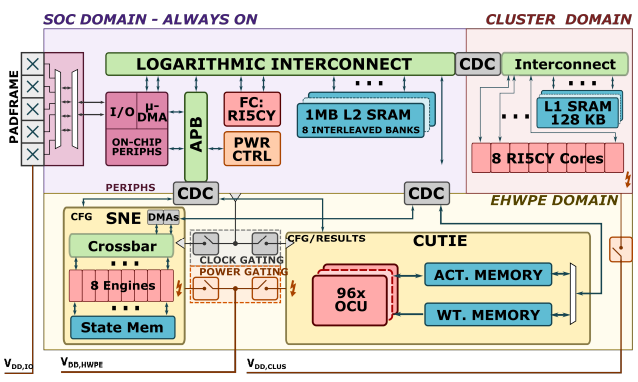}
    \caption{High-level architecture of Kraken SoC.}
    \label{fig:kraken}
\end{figure}

\textbf{\emph{Kraken}, SoC with SNN and ANN accelerators}: 
Kraken~\cite{dimauro2022kraken} (Figure~\ref{fig:kraken}) is an example for an ultra-low-power heterogeneous SoC fabricated in \SI{22}{\nano\metre} and combines a 32-bit RISC-V host core, \SI{1}{\mebi\byte} of scratchpad L2 SRAM memory, and an autonomous I/O subsystem with three programmable, power-gateable accelerators:
(1) A \SI{1.8}{\tera\op/\second/\watt} parallel general-purpose compute cluster with 8 RISC-V cores sharing \SI{128}{\kibi\byte} of L1 scratchpad memory. The RISC-V cores support hardware loops, SIMD sub-byte dot-product integer operations with mixed-precision capabilities, MAC with concurrent data load (MAC-LD), and floating-point capabilities for energy-efficient digital signal processing.
(2) \SI{1.1}{\tera\sy\op/\second/\watt} accelerator called \emph{Sparse Neural Engine (SNE)} targets spiking convolutional layers with 4-bit 3$\times$3 filter and 8-bit leaky-integrate and fire (LIF) neuron states.
(3) \emph{Completely Unrolled Ternary Inference Engine (CUTIE)}~\cite{scherer2022cutie} is a \SI{1036}{\tera\op/\second/\watt} Ternary Neural Network (TNN) accelerator. 

\textbf{\emph{$\mu$Brain}, Digital SNN} 
A fundamentally different approach to improving energy efficiency is one of \emph{neuromorphic} devices, which takes inspiration from the brain and research spiking neural networks both at the algorithmic and the hardware implementation fronts. A key difference between ANNs and SNNs is the stateful nature of spiking neurons, compared to the statefulness of the ReLU functions and the fact that SNNs communicate by passing a 1-bit message or spike, thus, resulting in sparse operation. Recent trends in neural network hardware are mainly three ($\mu$Brain die micrograph (left) and its fixed network architecture (right)~\cite{stuijt2021mubrain}) 1) CMOS-based neuromorphic SNN 2) traditional ANN accelerators 3) non-volatile-memory-based accelerators. 

\begin{figure}[t]
    \includegraphics[width=0.9\columnwidth]{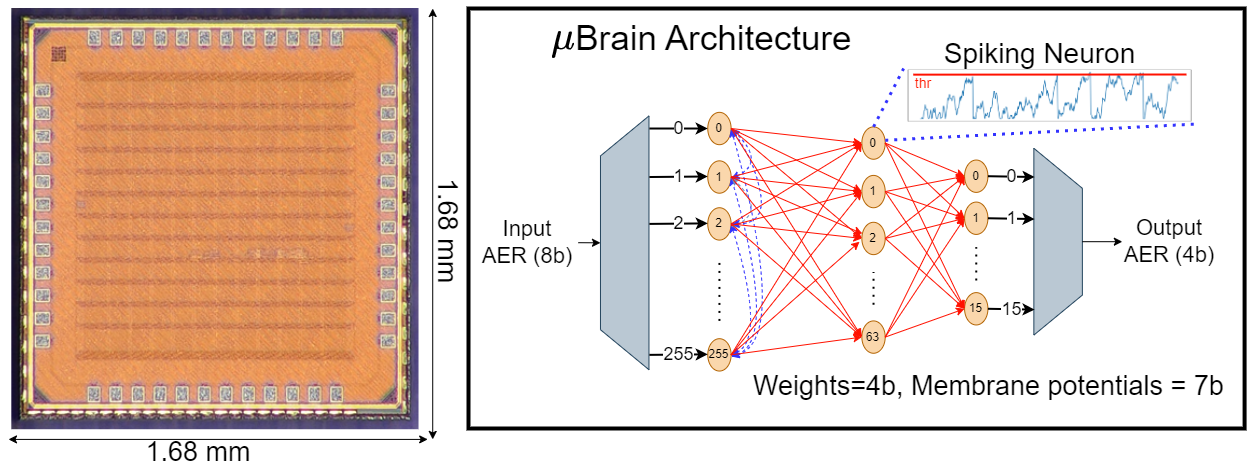}
    \caption{$\mu$Brain die micrograph (left) and its fixed network architecture (right)~\cite{stuijt2021mubrain}.}
    \label{fig:ubrain}
\end{figure}

\begin{figure}[t!]
\centering
    \includegraphics[width=0.9\columnwidth]{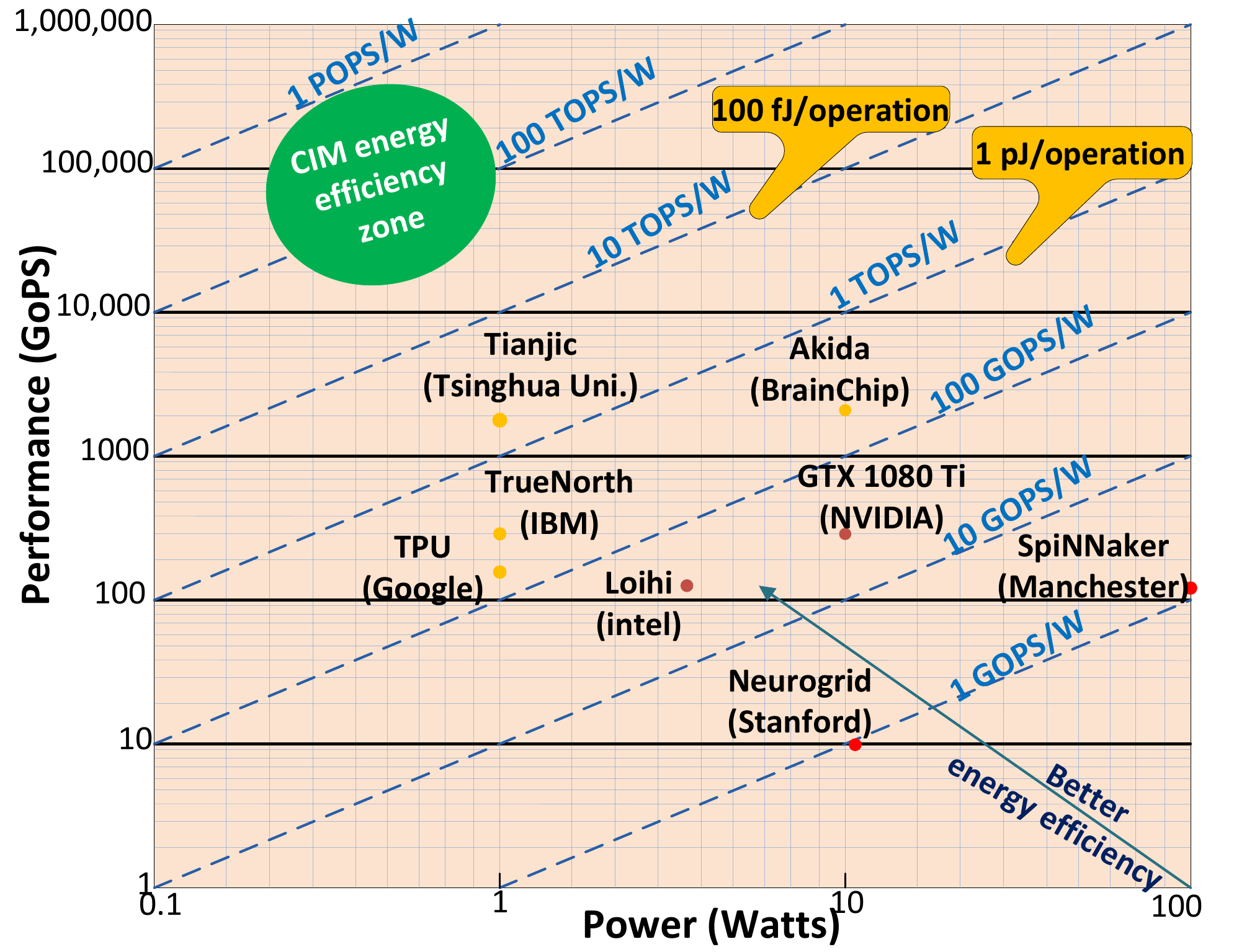}
    \caption{Energy-efficiency of various neuromorphic chips and CIM potentials~\cite{guo2017survey,gebregiorgis2022dealing}.}
    \label{fig:cimpot}
\end{figure}

\textbf{Summary:}
As shown in Figure~\ref{fig:cimpot} the energy-efficiency of existing neuromorphic chips is limited (few pJ/operation)~\cite{guo2017survey,gebregiorgis2022dealing,stuijt2021mubrain}.
\par

\section{CONVOLVE methodology}

CONVOLVE (\url{convolve.eu}) proposes a novel three-pillar design methodology, on which relies its four key objectives: \emph{(1)~Achieve 100X energy efficiency, (2)~Reduce design time by 10X, (3)~Guarantee hardware security and reliability and (4)~Enable smart edge applications}, as shown in Figure 7. The first pillar: \emph{ULP building blocks} focuses on exploitation of different hardware acceleration possibilities at microarchitecture, circuit  and device level; the second pillar \emph{Smart and dynamic application models} focuses on capturing the dynamic application behaviour efficiently; and the third pillar \emph{Compositional and fast design flow} efficiently bridges the first two pillars by generating system architecture and mapping of application models to the vast amount of hardware acceleration possibilities. Each pillar covers different levels of the design stack leading to various design choices, discussed in Sec. IV. 

\begin{figure}[b]
  \centering
   \includegraphics[width=0.7\textwidth, trim = 2.5cm 7.5cm 9cm 3cm, clip]{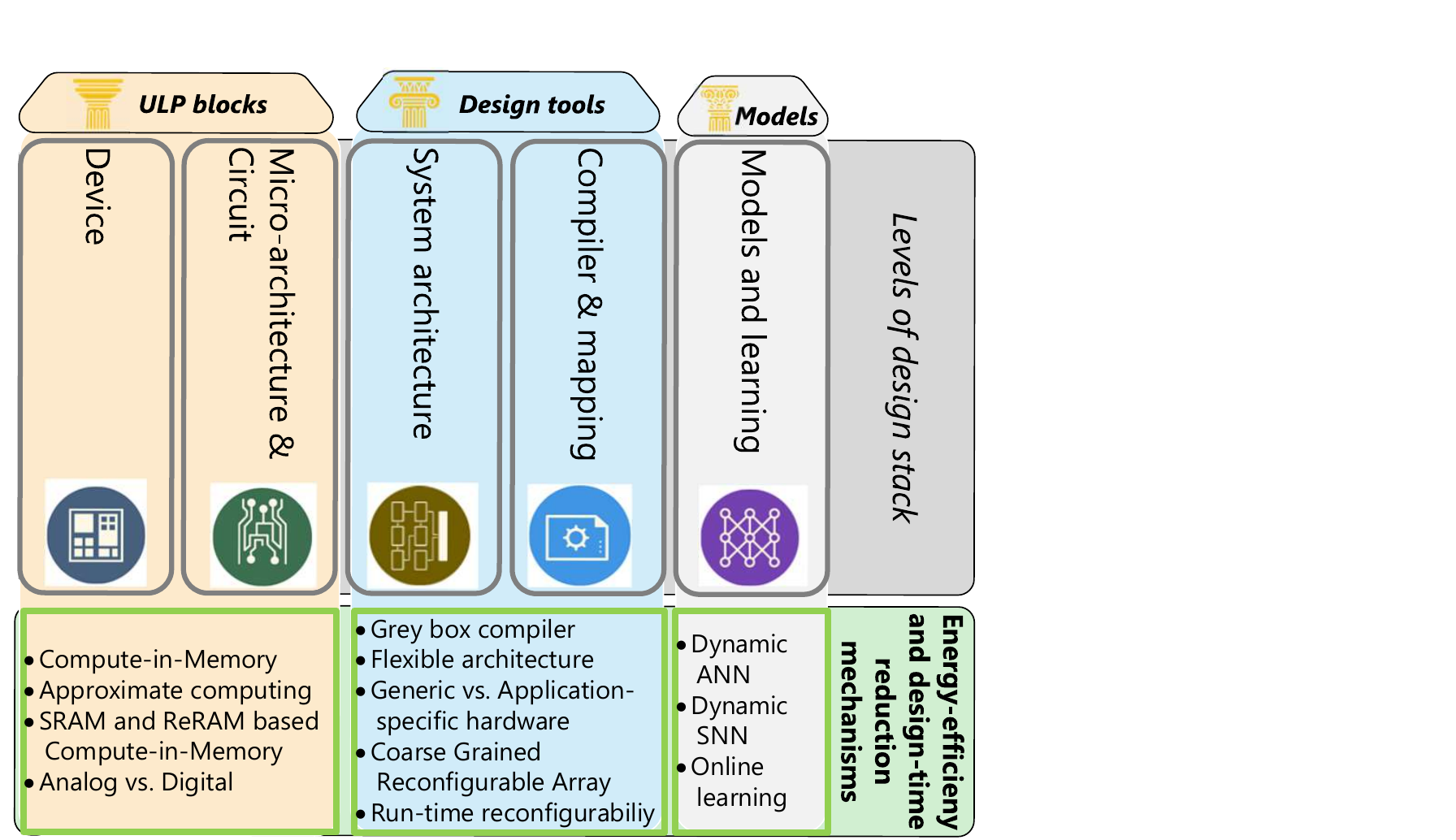} 
  \caption{Three-pillar design methodology of CONVOLVE to tackle energy efficiency and design time reduction objectives.}
  \label{Synapse Memory Architecture}
\end{figure}



\begin{figure}[t]
    \includegraphics[width=0.6\textwidth, trim = 6cm 11cm 4cm 1cm, clip]{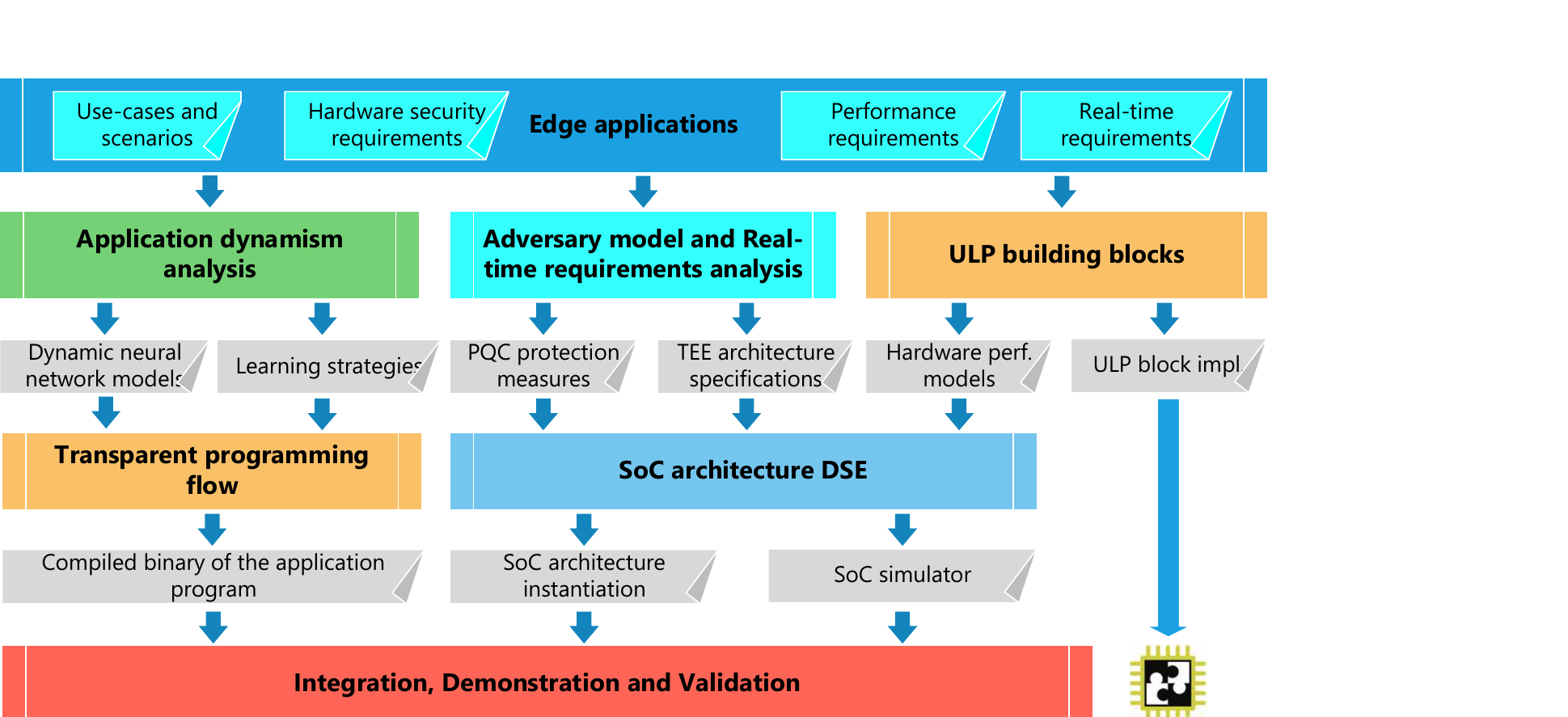} 
    \caption{CONVOLVE compositional and fast design flow: The different steps for the generation of fully integrated optimized SoC architecture from the application specification.}
    \label{fig:method}
\end{figure}

CONVOLVE's design flow starts from a given application suite, selects in a very short time the optimal SoC configuration, implements and verifies it, and compile algorithms for the generated hardware, as shown in Fig.~\ref{fig:method}. The goal is to automatically generate the optimal processing system for any given edge AI application, based on the ULP building blocks and their code generation, including the building blocks for hardware security. The application use-cases and scenarios are analyzed for understanding application dynamism that will define the dynamic neural network model and the learning strategies needed. An efficient, transparent and security-aware compilation flow built within the MLIR~\cite{lattner2020mlir} framework will be used to generate code for the heterogeneous set of ULP building blocks. MLIR is an industry-supported framework for optimizing programs by incremental lowering through a multitude of domain-specific IRs. A fully automated framework for DSE  and hardware generation will be based on the ZigZag ML performance estimation model~\cite{ZigZag}. The DSE framework uses as input performance models at two different levels: Core-level and SoC-level.



\textbf{Core-Level Modeling:} At the single-ULP-processing-core level, we model each accelerator’s architecture, including its run-time configurability, which enables rapid cost estimation for the design space of mapping a wide range of ML workloads onto each individual accelerator. In this way, the vast combinatorial space of hardware, algorithm, and mapping can be separately/jointly explored in a fast manner. This is beneficial for 1) hardware designers aiming at optimizing accelerators’ design-time/run-time architectural parameters, 2) algorithm developers working towards constructing hardware-friendly/compatible artificial neural networks’ topologies, and 3) compiler builders aspiring to design flexible compilation flows that can easily be adapted for evolving ML algorithms, as well as new ML accelerators, optimally mapping between the two.

\textbf{SoC-Level Modeling:} New trends of larger and more varied ML workloads require more performant and flexible hardware platforms. Both homogeneous and heterogeneous multi-core/-accelerator SoCs for ML are becoming ever more popular in recent years. At the SoC level, we will model and estimate the cost of end-to-end mapping one or multiple ML workloads onto the SoC system. This requires modeling not only each core/accelerator’s own attributes and mapping each operator of an ML algorithm one-at-a-time, but also the interconnection between the cores/accelerators and fine-grained data dependencies between operators. In this way, early-phase design space exploration of graph lowering and optimization, workload-core allocation, and inter-/intra-core scheduling can be performed in parallel with workload, HW and compiler development, and provide early feedback to other stages.

\section{CONVOLVE design space}

\subsection{Pillar: Smart and dynamic application models}

\textbf{ANN vs SNN + Online learning vs Offline learning} 
Many edge applications require constant monitoring of data streams on a tiny power budget. Examples include acoustic scene analysis, speech denoising, and keyword spotting. Furthermore, most current solutions rely on \acp{RNN} that process their input frame by frame. 
This frame-based approach is inefficient for these applications because it does not use sparsity in the input.
The limited computational capabilities of low-power edge devices require much of the processing in the cloud. 
However, non-local cloud processing results in increased energy costs and latency due to data movement. 
Finally, additional security and data privacy risks may be inherent to information transmission that could remain local provided sufficient compute power to process it on-device.
These limitations call for innovating toward smart dynamic application models.

Compared to \acp{RNN}, brain-inspired \acp{SNN} are still a relatively immature technology. However, they may offer compelling long-term solution to above problems. In \acp{SNN}, neurons communicate through stereotypical events, so-called spikes that are often sparse in time. This sparsity of activity is also referred to as  \textit{ephemeral sparsity}~\cite{hoefler_sparsity_2021}. The sparseness of activity increases the information content of each message passing between neurons, promising more energy-efficient computation and communication: 
When there is no input, no spikes propagate through the network, reducing memory access and thus saving energy. This event-driven processing model used by the brain may be a better fit for the continuous processing of sparse real-world data.

However, \acp{SNN} are not a mature technology, and much remains to be done. There are no well-established learning strategies to rival the success of \ac{BPTT} in conventional \acp{RNN}. Until recently, the inability to readily define a differentiable error function for spiking neurons posed a major conceptual obstacle. Today, new methods sidestep this problem by introducing practical surrogate gradients~\cite{neftci2019surrogate}. Combined with other innovations, \acp{SNN} can now be optimized without \ac{BPTT}~\cite{kaiser_synaptic_2020, yin2021accurateonline, zenke_brain-inspired_2021}, thereby enabling end-to-end training on otherwise prohibitively long sequences and paving the way for real-time on-chip learning. Combined with bio-inspired neuronal mechanisms such as heterogeneous adaption \cite{yin2021accurate,perez-nieves_neural_2021}, \ac{SNN} task performance is becoming increasingly competitive with \acp{RNN}. Capitalizing on the merits of this active research area will be a significant focus of CONVOLVE.

Beyond the sparsity of network activity, the connectivity between layers of neurons can itself be sparse, further reducing the memory footprint required to store the connectivity matrix and impacting silicon area and system cost as well as the energy cost of moving data. This structural sparsity can also be the target of biologically-inspired learning rules, termed \textit{structural plasticity}, permitting the creation and destruction of inter-neural connections. This is again a significant area of research in the project. From an implementation point of view, managing sparse connectivity matrices is challenging. Much of the performance gains in recent hardware for neural networks have come from parallelism, typified by the \ac{GPU} in which multi-lane compute resources are kept occupied with contiguous data fed by wide memory buses. Sparse matrices violate many of the underlying assumptions of such architectures, which must be addressed to avoid crippling inefficiencies in the resulting hardware. 
To support sparse connectivity efficiently in hardware is a major area for future innovations in the CONVOLVE project that emphasizes on the co-design of models and accelerators and seeks opportunities to combine the best of the \ac{ANN} and \ac{SNN} worlds. 

Finally, a key CONVOLVE focus area is improving the ability to learn continuously during deployment. Such online continual learning is essential for edge-based systems allowing them to seamlessly adapt to different users, environments, or task requirements. Similar online adaptation may also bestow self-healing capabilities to the system by providing robustness to component or sensor drift over the system’s lifetime. Solving these challenges requires real-time on-chip learning capabilities robust to catastrophic forgetting. However, on-chip learning precludes using \ac{BPTT} and supervised learning, thus requiring major algorithmic innovations. To address these points, within CONVOLVE we will develop new algorithms for self-supervised learning in continuous time that dispense with or at least minimize back-propagation requirements through time and space.




\textbf{Static vs Dynamic NN}
Several neural network models have been developed in recent years and have demonstrated excellent performance in many application domains. However, due to their computational complexity, these models may not be suitable for low-resource devices or latency-critical applications. Typical approaches to reducing the processing complexity of these models include model compression and response approximation. They aim at reducing the model size by injecting sparsity, adding collaborative layers \cite{Lee2021} or designing tiny architectures \cite{Liberis2021}. Dynamic neural networks (DNN) were introduced to make the processing complexity at the inference stage input-dependent. The idea behind DNN is borrowed from biological neural networks which are believed to adapt the neural pathways to the stimulus in order to speed up decision-making \cite{Ariely2011}. 

The most straightforward implementation of DNN is through Early Exit \cite{Scardapane2020}. It involves using internal classifiers to make quick decisions for easy inputs, i.e. without using the full-fledged network. A response is returned if the internal classifier is sufficiently confident; otherwise, the example is passed on to a subsequent internal classifier. Other studies made input dependence possible through: attention mechanisms which allow focusing on the most important parts of the input data \cite{Hu2018}; gating functions that remove the least salient components (eg. channels of an image) \cite{Gao2018}; parameter adaptation that aims at adaptively generating or altering the architecture’s intrinsic characteristics (eg. network width or depth) given the input’s features \cite{Xia2022}; and dynamic activation functions that activate neurons according to the relevance of the input stimulus, thus increasing the representation power of models \cite{Chen2020}. A comprehensive review of DNN can be found in \cite{Han2022}.


We aim to develop efficient DNN for smart low-resource processors. Various constraints will be applied to the models, including those related to computational complexity, latency, energy consumption, and performance (e.g. precision, robustness). We will  first review existing DNN from a resource-limited setting viewpoint, investigate the multi-criterion neural architecture search paradigm, and explore recent compression techniques (e.g. pruning-at-initialisation \cite{Wang2022}). Both supervised and self-supervised learning approaches will be investigated to train DNN models. 

\begin{figure}
\includegraphics[width=\columnwidth]{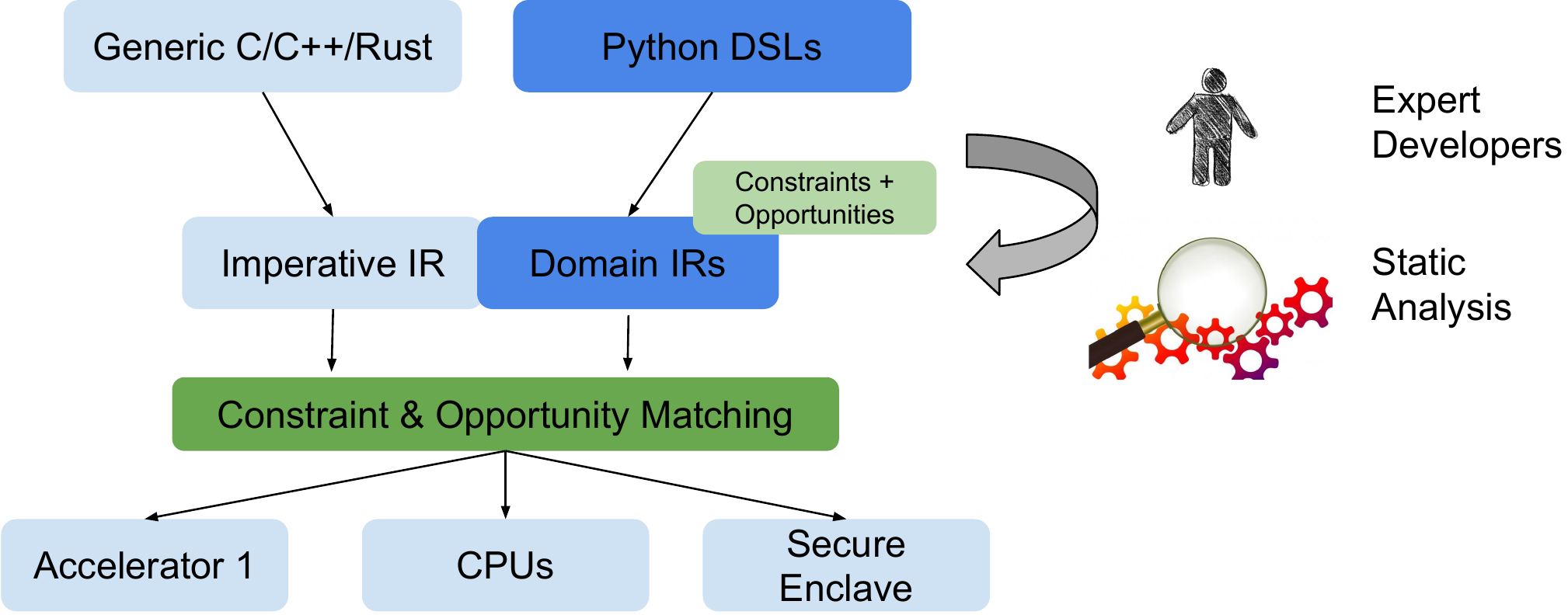}
\caption{The CONVOLVE gray-box compiler established a semi-automatic
compilation flow where static-analyses and expert developers collaborate
to obtain peak-performance when running large neural networks on the
CONVOLVE accelerators.}
\label{fig:compiler}
\end{figure}

\subsection{Pillar: Architecture and design tools}
\textbf{Black-box vs Grey-box compiler} To effectively map complex neural networks to the heterogeneous CONVOLVE
hardware, our compiler must scale to large applications while ensuring a code
quality that matches handwritten kernels developed by domain-expert programmers. While typical black-box compilers such
as LLVM~\cite{lattner2004llvm} offer a generic performance baseline, neural
networks are increasingly targeted via domain-specific frameworks such
as MLIR~\cite{lattner2020mlir} or TVM~\cite{chen2018tvm}. These can significantly improve performance and targetability by rigidly building in domain knowledge within a fixed-function stack, but just as the overspecialization in the frontend of LLVM hurts targetability of accelerators, the overspecialization in the backend of these newer frameworks hurts targetability of new applications, or the intended applications on accelerators with altered capabilities (e.g. lower precision). The CONVOLVE compiler
(Figure~\ref{fig:compiler}) will extend MLIR to offer a generic \textit{grey}-box approach, where
a novel theory of \textit{constraints} and \textit{opportunities} will guide static analysers and
expert developers to symbiotically work towards optimal hardware mappings in support
of the fast-evolving CONVOLVE hardware ecosystem, by embedding knowledge throughout program transformation, and preserving key invariants.

Powerful static analyses of the applications, feeding information throughout the intermediate-representation stack, will enable our compiler to
semi-automatically target CONVOLVE accelerators. Based on the
results of the static analysis, the \textit{opportunities} exposed by the applications'
characteristics, and the \textit{constraints} imposed by the execution environment, the
compiler will attempt to customize the application and automatically generate
efficient code, optimized and tailored for the target architecture. 
In
addition, we will offer domain-expert developers efficient access to the compiler
internals to specialize code optimization and application-to-accelerator
mapping for each use case.
 
The compiler first analyzes the \textit{opportunities} exposed by the application, such
as the degree of parallelism, reduced precision, code layout, algorithmic
structure, or a limited input domain. Then it considers the \textit{constraints} of the
execution context: latency requirements, timing and security guarantees, the
accuracy of the target, performance penalties of a complex control flow, etc.
These will serve as a description of the applications and available hardware
platforms, preserved under transformation, to guide the compiler in optimizing the applications. We envision
optimizations such as remapping data in memory (e.g., data transfer
reduction), targeting fixed-function accelerators via algorithmic matching, and
reorganizing code layout to alleviate bandwidth bottlenecks (for new custom
hardware and for providing real-time constraint guarantees). CONVOLVE will
complement static analysis with the manual and test-oriented generation of
opportunities, verified dynamically or with profile guidance, that offers
domain experts the opportunity to guide our gray-box compiler and attain
peak performance.

\textbf{Static vs dynamic architecture - Design hierarchy level: Architecture} 
Dynamic Neural Networks based on ANN and SNN are a fast-moving field of research in terms of network design and optimization. Optimization knobs include the quantization strategy and granularity, the number and types of network layers, and the supported types of network dynamism. As such, the traditional ways of designing heterogeneous multi-core SoCs with different accelerators to maximize energy-efficiency might be less effective when the number of different configurations/accelerators explodes or the computation/network changes in the future. 

A Coarse-Grained Reconfigurable Architecture (CGRA) could be a good compromise between the flexibility of a general-purpose processor and the energy-efficiency of a fixed-function accelerator. A CGRA consists of a grid (array) of processing elements or functional units (FUs) that are interconnected through a (reconfigurable) switching fabric or network-on-chip (NoC). Similar to FPGAs, there are (coarse-grained) specialized DSP blocks and local memories (LM) to increase the energy-efficiency, but the reconfiguration overhead in CGRAs is much lower over FPGAs. The class of CGRAs covers a wide range of reconfigurable architectures, each with a different degree of programmability. A recent overview is provided in \cite{7818353}.

\begin{figure}[t]
    \includegraphics[width=\columnwidth]{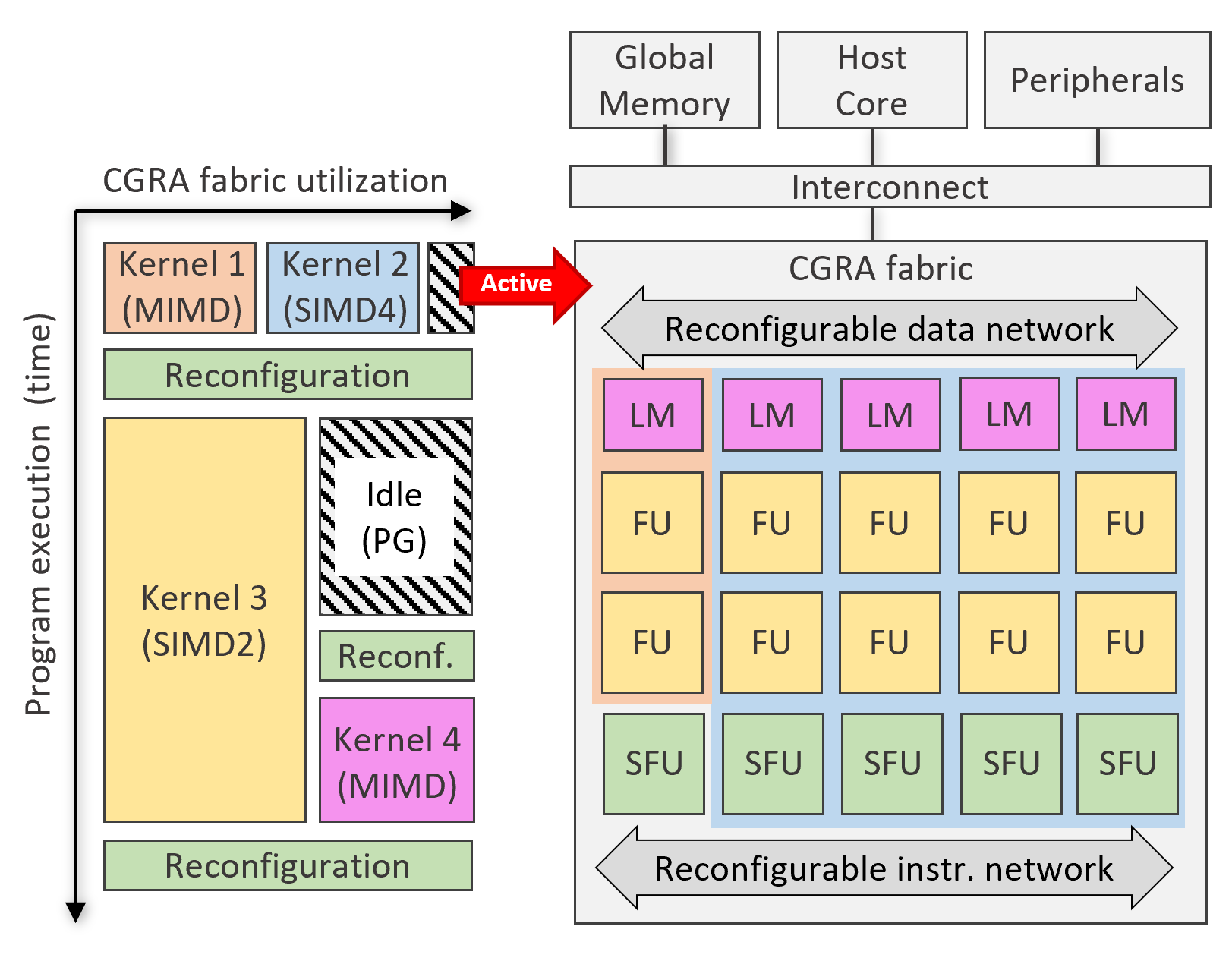}
    \caption{CONVOLVE CGRA fabric that executes multiple kernels in parallel while reconfiguring the fabric on a kernel-level granularity and power-gating unused units.}
    \label{fig:blocks_cgra}
\end{figure}

The CONVOLVE CGRA will be based on the Blocks CGRA template\cite{wijtvliet2021blocks} which will be extended for Dynamic Neural Network acceleration. An overview is depicted in Fig. \ref{fig:blocks_cgra}. The Blocks CGRA template consists of a reconfigurable instruction and data network with programmable FUs. The physical fabric (physArch) is able to execute multiple application-specific processors or virtual cores (virtArchs) in parallel. These virtual cores can be VLIW cores with optional SIMD extensions sized towards the kernel or application at hand. The fabric supports flexible SIMD execution by broadcasting the same instructions to multiple FUs over the reconfigurable instruction network. Complex instructions are supported via specialized FUs (SFUs). To accommodate efficient acceleration for varying workloads, the fabric can be (partially) reconfigured on a kernel-level granularity with a tolerable reconfiguration penalty\cite{9580639}. The fabric contains support for zero-overhead loops to enable spatial computation where the instruction stream remains static. Leakage power of unused FUs can be reduced using power gating.

To summarize, the use of a CGRA in CONVOLVE has several advantages: (1) it supports highly parallel calculations, (2) it has good area-efficiency compared to an FPGA, (3) it has high energy-efficiency due to the static interconnect and spatial mapping of computation where possible, (4) and it is flexible, supporting all kinds of applications as long as the FUs are not too specialized. Future research is needed within the CONVOLVE project to optimize the CGRA fabric for Dynamic Neural Network acceleration and to improve the existing retargetable C-compiler for efficient code generation with support for run-time (partial) reconfiguration.

\textbf{Generic vs Application specific hardware - Design hierarchy level: Architecture}
Most commercially available computing platforms are based on traditional general-purpose CPUs, which offer full flexibility and easy programmability. Such architectures are suitable for a variety of applications and hence allow mass-deployment and increased production volume which in the end helps to reduce the fabrication cost per unit. However, being general-purpose often results in mediocre performance and sub-optimal energy efficiency.
In contrast, application-specific accelerators can achieve optimal throughput, area and energy efficiency for a specific hardwired algorithm. Such dedicated accelerators have very limited capabilities to adapt to next-generation algorithms, which --- in the worst case --- could lead to the expensive dedicated silicon area becoming completely unusable and requiring an expensive re-design of the accelerator.
A balance between these two extremes is offered by FPGA solutions, where the datapath can be rewired to adapt to novel algorithms. However, FPGA platforms are still rather expensive when targeting a massive and dense deployment and the SoC integration is still rather difficult: Embedded FPGA solutions (e.g., offered by Menta, QuickLogic) which are offered as soft IPs are slower and more area-hungry than dedicated FPGAs chips.
Another option is offered by general-purpose accelerators such as \acp{GPU}. Their flexibility allows their use for various applications in a variety of scenarios and hence offers a more cost-efficient alternative than FPGA or application-specific accelerators. Nevertheless, their somewhat hardwired datapath can limit accelerator utilization and hence deliver sub-optimal performance for certain kernels.
The in CONVOLVE developed ULP acceleration blocks target a trade-off between application-specific and general-purpose computing by combining specialization with modularity, dynamic reconfigurability, and self-healing capabilities to fully harness the potential of machine learning. Furthermore, the selection of reconfigurable ULP blocks plan to support dynamic neural networks, improve reliability against process variations and provide real-time guarantees for safety-critical applications, and have provision for hardware security against e.g., side-channel attacks and security futureproofing.
To keep the energy-efficiency high despite these reconfiguration overheads, CONVOLVE  explores novel circuit and architecture-level techniques for efficient SRAM/ReRAM based CIM and CGRA.

\subsection{Pillar: ULP blocks}
\textbf{In-memory vs Classical computing architecture- Design hierarchy level - circuit/architecture/device} 
Moore’s law has enabled the traditional von Neumann-based CPUs to deliver better performance for successive generations~\cite{rai2021perspectives}. However, traditional CPU architectures are facing three major walls: (1) the memory wall due to the growing gap between processor and memory speed, and the limited memory bandwidth; (2) the Instruction-Level parallelism (ILP) wall due to the difficulty of extracting sufficient parallelism to fully exploit all the cores; (3) the power wall as the CPU clock frequency has reached the practical maximum value that is limited by cooling. In order for computing systems to continue delivering the required performance given the economical power constraints, novel computer architectures in the light of emerging non-volatile (practically no leakage) device technologies have to be explored. CIM has the potential to overcome those challenges by integrating computation and storage of data within the same physical location~\cite{singh2022cim,singh2021low}. CIM can be realized using different emerging memristive technologies such as Resistive Random Access Memory (RRAM), Phase Changing Memory (PCM) and Magnetic RAM (MRAM) as well as conventional memory technologies such as SRAM, DRAM and Ferroelectric FETs~\cite{ singh2021low,si2019twin}. In-memory computing using emerging memrisitve devices benefits from their non-volatile nature and their practically zero leakage compared to their conventional memory technology counterparts. Table~\ref{tablmemories} presents qualitative comparison of traditional von-Neumann based CPU and CIM architectures using different memory technologies~\cite{salahuddin2018era,oboril2015evaluaion}. CONVOLVE explores ultra-low power implementation of domain specific analog and digital CIM flavors using different memory technologies as well as Coarse-Grained Reconfigurable Arrays (CGRA).\\ 
\begin{table}[!t]
\caption{Design metrics for von-Neumann and CIM architectures using various memory technologies (data obtained from~\cite{salahuddin2018era,oboril2015evaluaion})}
\resizebox{\linewidth}{!}{%
 \begin{tabular}{|c|c|c|c|c|}
		\hline
		\begin{tabular}[]{@{}c@{}}Comparison\\ metric\end{tabular}   & \begin{tabular}[]{@{}c@{}}Conventional\\ CPU\end{tabular} & \begin{tabular}[c]{@{}c@{}}CIM digital\\ SRAM\end{tabular} & \begin{tabular}[c]{@{}c@{}}CIM analog \\ SRAM \end{tabular} & \begin{tabular}[c]{@{}c@{}}CIM analog\\ memristive \end{tabular} \\ 
		\hline
		\begin{tabular}[]{@{}c@{}}Device \\technology\end{tabular} &  \cellcolor{red!25} CMOS & \cellcolor{red!25} CMOS & \cellcolor{red!25} CMOS & \cellcolor{green!25} RRAM \\ 
		\hline
		Architecture & \cellcolor{red!25} von-Neumann & \cellcolor{red!25} von-Neumann & \cellcolor{green!25} \begin{tabular}[]{@{}c@{}} non \\ von-Neumann\end{tabular} & \cellcolor{green!25} \begin{tabular}[]{@{}c@{}} non \\ von-Neumann\end{tabular}\\
		\hline
		\begin{tabular}[]{@{}c@{}} Mode of \\ operation\end{tabular} &  \cellcolor{green!25} Digital & \cellcolor{green!25} Digital & \cellcolor{yellow!25} Analog & \cellcolor{yellow!25} Analog\\
		\hline
		Volatility & \cellcolor{red!25} Yes & \cellcolor{red!25} Yes & \cellcolor{red!25} Yes & \cellcolor{green!25} No \\ 
		\hline
		Endurance & \cellcolor{green!25} High & \cellcolor{green!25} High & \cellcolor{green!25} High & \cellcolor{red!25} Low\\
		\hline
		Scalability & \cellcolor{yellow!25} medium & \cellcolor{yellow!25} medium & \cellcolor{yellow!25} medium & \cellcolor{green!25} high\\ 
		\hline
		Write energy & \cellcolor{green!25} $\sim$fJ & \cellcolor{green!25} $\sim$fJ & \cellcolor{green!25} $\sim$fJ & \cellcolor{yellow!25} $\sim$pJ \\ 
		\hline
		Write latency & \cellcolor{green!25} $\sim$1ns & \cellcolor{green!25} $\sim$1ns & \cellcolor{green!25} 1ns & \cellcolor{yellow!25} $\sim$10ns\\ 
		\hline
		Read latency & \cellcolor{green!25} $\sim$1ns & \cellcolor{green!25} $\sim$1ns & \cellcolor{green!25} $\sim$1ns & \cellcolor{yellow!25} $\sim$10ns\\ 
		\hline
	\end{tabular}}
\label{tablmemories}
\end{table}

\textbf{Exact vs Approximate computation - Design hierarchy level: Circuit} 
Diminishing energy-efficiency gains from semiconductor scaling as per Moore's law and continued increase in compute-requirements, as evident from latest machine learning (ML) models like GPT3, Transformers etc., has forced researchers to look for newer computing paradigms. \textit{Approximate Computing} (AxC), which trades off accuracy for improved energy-efficiency, emerges as a potential alternative owing to the error-resilient characteristics of modern ML workloads. 

While AxC techniques have shown benefits at all levels of the computing stack, zooming in on the circuit-level, most AxC techniques can be classified into three broad categories: (a) timing approximation, wherein the circuit is operated at a lower supply voltage without reducing corresponding operational frequency, resulting in efficiency improvements for added timing errors \cite{voltage_overscale}, (b) functional approximation, wherein the logic functionality of the circuit is modified to trade off quality for added efficiency e.g. netlist modification \cite{netlist_mod}, boolean rewriting \cite{salsa}, \cite{dacLXSP19}, precision-scaling \cite{bfloat, dlfloat, ternary, choi2018pact} etc. and c) cross-level approximation where approximation knobs at both logic-, structural- and physical level are leveraged in a disciplined manner to boost energy efficiency \cite{tcasZXSP19} have been implemented for functional approximation.

AxC has been widely adopted for ML. Among all different approximations being investigated for ML, \textit{precision-scaling} has emerged as a success story. ML models have been shown to tolerate very aggressive precision scaling. It provides very high gains by reducing both compute as well as off-chip traffic in memory. For training, different data formats like 16-bit floats (FP16), BFloat \cite{bfloat}, DLFloat \cite{dlfloat} etc. have been adopted for activation and weights. For inference, varying fixed-point formats are adopted scaling all the way down to ternary or binary bitwidths \cite{ternary}. Accuracy lost due to approximations are regained using quantization-aware training (QAT) \cite{choi2018pact}. Execution engines with varying precision support have been proposed in both academia and industry \cite{survey_ai_accl}. Another widely adopted approximation technique is \textit{pruning} which forces weight values in a neural network to zero, thereby introducing sparsity. Several studies have proposed the combination of weight pruning with precision scaling to achieve higher energy efficiency for ML inference \cite{lascasLMXPS22, 8942068}. Pruning introduces irregularities in compute and memory access pattern. To tackle such challenges, specialized architectures with sparsity support have been proposed in literature \cite{sparsity}. Motivated by the promising returns proposed by the close synergy of AxC and ML, CONVOLVE aims at exploring novel AxC techniques at the circuit level to obtain extremely energy-efficient edge AI microprocessors.

\textbf{Analog vs Digital computing - Design hierarchy level: Circuit/device}: Current ANN/SNN compute architectures can be classified into whether they can be implemented using fully synthesizable logic using standard-cell library (i.e., all-digital) and custom-designed using analog/mixed-signal design techniques. Although state-of-the-art analog and mixed-signal architectures offer the lowest energy consumption, they have severe limitations that limited their applications. Firstly, they offer poor reliability as their performance is easily affected by noise induced by process, temperature, and voltage variations and hence require a complex calibration process. Secondly, they offer poor technology portability as they need to be re-designed when porting the design to a different technology node. Thirdly, they offer poor scalability as larger designs cannot be easily built using powerful design automation tools available for digital designs~\cite{SchumanPPBDRP17, abs-2005-01467}. 
However, analog implementations are proven to be more energy-efficient compared to digital. More recently, Wan et al.~\cite{wan2022compute} demonstrate the exploitation of highly-integrated resistive random-access memory (ReRAM) devices to avoid power-hungry data movement between separate compute and memory, achieving inference accuracy comparable to software models trained with 4-bit weights across several AI benchmark tasks. These achievements demonstrate a simultaneous improvement in efficiency, flexibility, and accuracy over existing RRAM-CIM hardware by innovating across the entire design hierarchy, from a reconfigurable dataflow architecture to an energy- and area-efficient voltage-mode neuron circuit and to a series of algorithm-hardware co-optimization techniques. This demonstrates that the problem of efficient computing must be tackled simultaneously and at multiple levels of the stack.  




\section{Summary}
CONVOLVE aims at 100X improvement in energy efficiency and 10X in design-time. We outlined the SotA for edge-AI processing, the CONVOLVE automated design flow,
and treated  8 important design choices supporting above goals. Major conclusions are:
1) In the short term, ANNs are to be favored above SNNs; 
2) Although SNNs have favorable properties, inspired by our brain, they require more research;
3) Online learning requires rethinking back-propagation through time and space; 
4) Dynamic \acp{NN} will become dominant in low-power edge computing;
5) Quick and easy adaptation requires a Grey-box compiler that can deal with new accelerators comprehensively;
5) Architectures should support more dynamism, e.g., by using 2-step code generation (i.e., Code $\rightarrow$ Virtual cores $\rightarrow$ Physical architecture);
6) The ML field is moving quickly and therefore requires flexible architectures;
7) CIM is promising but requires adapting the memory periphery;
8) Analog computing has a high potential for energy savings but is too inflexible in the short term; since \acp{NN} do not fit spatially yet,  time-sharing and reconfiguration flexibility are key.
CONVOLVE aims at unleasing above potential by a holistic approach, rethinking the full design stack. 



\bibliographystyle{IEEEtran}
\bibliography{ref_control,references} 

\begin{thebibliography}{10}
\providecommand{\url}[1]{#1}
\csname url@samestyle\endcsname
\providecommand{\newblock}{\relax}
\providecommand{\bibinfo}[2]{#2}
\providecommand{\BIBentrySTDinterwordspacing}{\spaceskip=0pt\relax}
\providecommand{\BIBentryALTinterwordstretchfactor}{4}
\providecommand{\BIBentryALTinterwordspacing}{\spaceskip=\fontdimen2\font plus
\BIBentryALTinterwordstretchfactor\fontdimen3\font minus
  \fontdimen4\font\relax}
\providecommand{\BIBforeignlanguage}[2]{{%
\expandafter\ifx\csname l@#1\endcsname\relax
\typeout{** WARNING: IEEEtran.bst: No hyphenation pattern has been}%
\typeout{** loaded for the language `#1'. Using the pattern for}%
\typeout{** the default language instead.}%
\else
\language=\csname l@#1\endcsname
\fi
#2}}
\providecommand{\BIBdecl}{\relax}
\BIBdecl

\bibitem{tinyVers}
V.~Jain \emph{et~al.}, ``Tinyvers: A 0.8-17 tops/w, 1.7 uw-20 mw, tiny
  versatile system-on-chip with state-retentive emram for machine learning
  inference at the extreme edge,'' in \emph{2022 IEEE Symposium on VLSI
  Technology and Circuits (VLSI Technology and Circuits)}, 2022, pp. 20--21.

\bibitem{diana}
K.~Ueyoshi \emph{et~al.}, ``Diana: An end-to-end energy-efficient digital and
  analog hybrid neural network soc,'' in \emph{2022 IEEE International Solid-
  State Circuits Conference (ISSCC)}, vol.~65, 2022, pp. 1--3.

\bibitem{si2019twin}
X.~Si \emph{et~al.}, ``A twin-8t sram computation-in-memory unit-macro for
  multibit cnn-based ai edge processors,'' \emph{IEEE Journal of Solid-State
  Circuits}, vol.~55, no.~1, pp. 189--202, 2019.

\bibitem{9731754}
H.~Fujiwara \emph{et~al.}, ``A 5-nm 254-tops/w 221-tops/mm2 fully-digital
  computing-in-memory macro supporting wide-range dynamic-voltage-frequency
  scaling and simultaneous mac and write operations,'' in \emph{2022 IEEE
  International Solid- State Circuits Conference (ISSCC)}, vol.~65, 2022, pp.
  1--3.

\bibitem{roy2019towards}
K.~Roy \emph{et~al.}, ``Towards spike-based machine intelligence with
  neuromorphic computing,'' \emph{Nature}, vol. 575, no. 7784, pp. 607--617,
  2019.

\bibitem{singh2021srif}
A.~Singh \emph{et~al.}, ``Srif: Scalable and reliable integrate and fire
  circuit adc for memristor-based cim architectures,'' \emph{IEEE Transactions
  on Circuits and Systems I: Regular Papers}, vol.~68, no.~5, pp. 1917--1930,
  2021.

\bibitem{singh2021low}
------, ``Low-power memristor-based computing for edge-ai applications,'' in
  \emph{2021 IEEE International Symposium on Circuits and Systems
  (ISCAS)}.\hskip 1em plus 0.5em minus 0.4em\relax IEEE, 2021, pp. 1--5.

\bibitem{new2021unbalanced}
S.~Diware \emph{et~al.}, ``Unbalanced bit-slicing scheme for accurate
  memristor-based neural network architecture,'' in \emph{2021 IEEE 3rd
  International Conference on Artificial Intelligence Circuits and Systems
  (AICAS)}.\hskip 1em plus 0.5em minus 0.4em\relax IEEE, 2021, pp. 1--4.

\bibitem{bengel2022reliability}
C.~Bengel \emph{et~al.}, ``Reliability aspects of binary
  vector-matrix-multiplications using reram devices,'' \emph{Neuromorphic
  Computing and Engineering}, 2022.

\bibitem{mayahinia2022voltage}
M.~Mayahinia \emph{et~al.}, ``A voltage-controlled, oscillation-based adc
  design for computation-in-memory architectures using emerging rerams,''
  \emph{ACM Journal on Emerging Technologies in Computing Systems (JETC)},
  vol.~18, no.~2, pp. 1--25, 2022.

\bibitem{diware2022accurate}
S.~Diware \emph{et~al.}, ``Accurate and energy-efficient bit-slicing for
  rram-based neural networks,'' \emph{IEEE Transactions on Emerging Topics in
  Computational Intelligence}, 2022.

\bibitem{spetalnick202240nm}
S.~D. Spetalnick \emph{et~al.}, ``A 40nm 64kb 26.56 tops/w 2.37 mb/mm 2 rram
  binary/compute-in-memory macro with 4.23 x improvement in density and> 75\%
  use of sensing dynamic range,'' in \emph{2022 IEEE International Solid-State
  Circuits Conference (ISSCC)}, vol.~65.\hskip 1em plus 0.5em minus 0.4em\relax
  IEEE, 2022, pp. 1--3.

\bibitem{wan2022compute}
W.~Wan \emph{et~al.}, ``A compute-in-memory chip based on resistive
  random-access memory,'' \emph{Nature}, vol. 608, no. 7923, pp. 504--512,
  2022.

\bibitem{xue202116}
C.-X. Xue \emph{et~al.}, ``16.1 a 22nm 4mb 8b-precision reram
  computing-in-memory macro with 11.91 to 195.7 tops/w for tiny ai edge
  devices,'' in \emph{2021 IEEE International Solid-State Circuits Conference
  (ISSCC)}, vol.~64.\hskip 1em plus 0.5em minus 0.4em\relax IEEE, 2021, pp.
  245--247.

\bibitem{dimauro2022kraken}
A.~Di~Mauro \emph{et~al.}, ``Kraken: A direct event/frame-based multi-sensor
  fusion soc for ultra-efficient visual processing in nano-uavs,'' in
  \emph{2022 IEEE Hot Chips 34 Symposium (HCS)}, 2022, pp. 1--19.

\bibitem{scherer2022cutie}
M.~Scherer \emph{et~al.}, ``A 1036 top/s/w, 12.2 mw, 2.72 $\mu$j/inference all
  digital tnn accelerator in 22 nm fdx technology for tinyml applications,'' in
  \emph{2022 IEEE Symposium in Low-Power and High-Speed Chips (COOL
  CHIPS)}.\hskip 1em plus 0.5em minus 0.4em\relax IEEE, 2022, pp. 1--3.

\bibitem{stuijt2021mubrain}
J.~Stuijt \emph{et~al.}, ``$\mu$brain: An event-driven and fully synthesizable
  architecture for spiking neural networks,'' \emph{Frontiers in neuroscience},
  vol.~15, p. 538, 2021.

\bibitem{guo2017survey}
K.~Guo \emph{et~al.}, ``A survey of fpga-based neural network accelerator,''
  \emph{arXiv preprint arXiv:1712.08934}, 2017.

\bibitem{gebregiorgis2022dealing}
A.~Gebregiorgis \emph{et~al.}, ``Dealing with non-idealities in memristor based
  computation-in-memory designs,'' in \emph{2022 IFIP/IEEE 30th International
  Conference on Very Large Scale Integration (VLSI-SoC)}.\hskip 1em plus 0.5em
  minus 0.4em\relax IEEE, 2022, pp. 1--6.

\bibitem{lattner2020mlir}
C.~Lattner \emph{et~al.}, ``Mlir: A compiler infrastructure for the end of
  moore's law,'' \emph{arXiv preprint arXiv:2002.11054}, 2020.

\bibitem{ZigZag}
L.~Mei \emph{et~al.}, ``{ZigZag}: Enlarging joint architecture-mapping design
  space exploration for dnn accelerators,'' \emph{IEEE Transactions on
  Computers}, vol.~70, no.~8, pp. 1160--1174, 2021.

\bibitem{hoefler_sparsity_2021}
T.~Hoefler \emph{et~al.}, ``Sparsity in {Deep} {Learning}: {Pruning} and growth
  for efficient inference and training in neural networks,''
  \emph{arXiv:2102.00554 [cs]}, Jan. 2021.

\bibitem{neftci2019surrogate}
E.~O. Neftci \emph{et~al.}, ``Surrogate gradient learning in spiking neural
  networks: Bringing the power of gradient-based optimization to spiking neural
  networks,'' \emph{IEEE Signal Processing Magazine}, vol.~36, no.~6, pp.
  51--63, 2019.

\bibitem{kaiser_synaptic_2020}
J.~Kaiser \emph{et~al.}, ``\BIBforeignlanguage{English}{Synaptic {Plasticity}
  {Dynamics} for {Deep} {Continuous} {Local} {Learning} ({DECOLLE})},''
  \emph{\BIBforeignlanguage{English}{Frontiers in Neuroscience}}, vol.~14,
  2020.

\bibitem{yin2021accurateonline}
B.~Yin \emph{et~al.}, ``Accurate online training of dynamical spiking neural
  networks through forward propagation through time,'' \emph{arXiv preprint
  arXiv:2112.11231}, 2021.

\bibitem{zenke_brain-inspired_2021}
F.~Zenke \emph{et~al.}, ``Brain-{Inspired} {Learning} on {Neuromorphic}
  {Substrates},'' \emph{Proceedings of the IEEE}, vol. 109, no.~5, pp.
  935--950, May 2021.

\bibitem{yin2021accurate}
B.~Yin \emph{et~al.}, ``Accurate and efficient time-domain classification with
  adaptive spiking recurrent neural networks,'' \emph{Nature Machine
  Intelligence}, vol.~3, no.~10, pp. 905--913, 2021.

\bibitem{perez-nieves_neural_2021}
N.~Perez-Nieves \emph{et~al.}, ``\BIBforeignlanguage{en}{Neural heterogeneity
  promotes robust learning},'' \emph{\BIBforeignlanguage{en}{bioRxiv}}, p.
  2020.12.18.423468, Jan. 2021.

\bibitem{Lee2021}
J.~Lee \emph{et~al.}, ``Resource-efficient deep learning: A survey on model-,
  arithmetic-, and implementation-level techniques,'' \emph{arXiv.2112.15131},
  2021.

\bibitem{Liberis2021}
E.~Liberis \emph{et~al.}, ``$\mu$nas: Constrained neural architecture search
  for microcontrollers,'' in \emph{1st Workshop on Machine Learning and
  Systems}, 2021, pp. 70--79.

\bibitem{Ariely2011}
D.~Ariely \emph{et~al.}, ``From thinking too little to thinking too much: a
  continuum of decision making,'' \emph{Wiley Interdisciplinary Reviews:
  Cognitive Science}, vol.~2, no.~1, pp. 39--46, 2011.

\bibitem{Scardapane2020}
S.~Scardapane \emph{et~al.}, ``Why should we add early exits to neural
  networks?'' \emph{Cognitive Computing}, vol.~12, no.~5, pp. 954--966, 2020.

\bibitem{Hu2018}
J.~Hu \emph{et~al.}, ``queeze-and-excitation networks,'' in \emph{2018 IEEE/CVF
  Conference on Computer Vision and Pattern Recognition}, 2018, pp. 7132--7141.

\bibitem{Gao2018}
X.~Gao \emph{et~al.}, ``Dynamic channel pruning: Feature boosting and
  suppression,'' \emph{arXiv:1810.05331v2}, 2018.

\bibitem{Xia2022}
W.~Xia \emph{et~al.}, ``Fully dynamic inference with deep neural networks,''
  \emph{IEEE Transactions on Emerging Topics in Computing}, vol.~10, no.~2, pp.
  962--972, 2022.

\bibitem{Chen2020}
Y.~Chen \emph{et~al.}, ``Dynamic relu,'' in \emph{16th European Conference on
  Computer Vision (ECCV)}, 2020, p. 351–367.

\bibitem{Han2022}
Y.~Han \emph{et~al.}, ``Dynamic neural networks: A survey,'' \emph{IEEE
  Transactions on Pattern Analysis and Machine Intelligence}, vol.~44, no.~11,
  p. 7436–7456, 2022.

\bibitem{Wang2022}
H.~Wang \emph{et~al.}, ``Recent advances on neural network pruning at
  initialization,'' in \emph{Thirty-First International Joint Conference on
  Artificial Intelligence}, 2022, p. 5638–5645.

\bibitem{lattner2004llvm}
C.~Lattner \emph{et~al.}, ``Llvm: A compilation framework for lifelong program
  analysis \& transformation,'' in \emph{International Symposium on Code
  Generation and Optimization, 2004. CGO 2004.}\hskip 1em plus 0.5em minus
  0.4em\relax IEEE, 2004, pp. 75--86.

\bibitem{chen2018tvm}
T.~Chen \emph{et~al.}, ``{TVM}: An automated {End-to-End} optimizing compiler
  for deep learning,'' in \emph{13th USENIX Symposium on Operating Systems
  Design and Implementation (OSDI 18)}, 2018, pp. 578--594.

\bibitem{7818353}
M.~Wijtvliet \emph{et~al.}, ``Coarse grained reconfigurable architectures in
  the past 25 years: Overview and classification,'' in \emph{2016 International
  Conference on Embedded Computer Systems: Architectures, Modeling and
  Simulation (SAMOS)}, 2016, pp. 235--244.

\bibitem{wijtvliet2021blocks}
------, ``Blocks: Challenging simds and vliws with a reconfigurable
  architecture,'' \emph{IEEE Transactions on Computer-Aided Design of
  Integrated Circuits and Systems}, vol.~41, no.~9, pp. 2915--2928, 2021.

\bibitem{9580639}
B.~de~Bruin \emph{et~al.}, ``Multi-level optimization of an ultra-low power
  brainwave system for non-convulsive seizure detection,'' \emph{IEEE
  Transactions on Biomedical Circuits and Systems}, vol.~15, no.~5, pp.
  1107--1121, 2021.

\bibitem{rai2021perspectives}
S.~Rai \emph{et~al.}, ``Perspectives on emerging computation-in-memory
  paradigms,'' in \emph{2021 Design, Automation \& Test in Europe Conference \&
  Exhibition (DATE)}.\hskip 1em plus 0.5em minus 0.4em\relax IEEE, 2021, pp.
  1925--1934.

\bibitem{singh2022cim}
A.~Singh \emph{et~al.}, ``Cim-based robust logic accelerator using 28 nm
  stt-mram characterization chip tape-out,'' in \emph{2022 IEEE 4th
  International Conference on Artificial Intelligence Circuits and Systems
  (AICAS)}.\hskip 1em plus 0.5em minus 0.4em\relax IEEE, 2022, pp. 451--454.

\bibitem{salahuddin2018era}
S.~Salahuddin \emph{et~al.}, ``The era of hyper-scaling in electronics,''
  \emph{Nature Electronics}, vol.~1, no.~8, pp. 442--450, 2018.

\bibitem{oboril2015evaluaion}
F.~Oboril \emph{et~al.}, ``Evaluation of hybrid memory technologies using
  sot-mram for on-chip cache hierarchy,'' \emph{IEEE Transactions on
  Computer-Aided Design of Integrated Circuits and Systems}, vol.~34, no.~3,
  pp. 367--380, 2015.

\bibitem{voltage_overscale}
K.~Shi \emph{et~al.}, ``Datapath synthesis for overclocking: Online arithmetic
  for latency-accuracy trade-offs,'' in \emph{Proceedings of the 51st Annual
  Design Automation Conference}, ser. DAC '14.\hskip 1em plus 0.5em minus
  0.4em\relax New York, NY, USA: Association for Computing Machinery, 2014, p.
  1–6.

\bibitem{netlist_mod}
S.~De \emph{et~al.}, ``An automated approximation methodology for arithmetic
  circuits,'' in \emph{2019 IEEE/ACM International Symposium on Low Power
  Electronics and Design (ISLPED)}, 2019, pp. 1--6.

\bibitem{salsa}
S.~Venkataramani \emph{et~al.}, ``Salsa: Systematic logic synthesis of
  approximate circuits,'' in \emph{DAC Design Automation Conference 2012},
  2012, pp. 796--801.

\bibitem{dacLXSP19}
V.~Leon \emph{et~al.}, ``Cooperative arithmetic-aware approximation techniques
  for energy-efficient multipliers,'' in \emph{Proceedings of the 56th Annual
  Design Automation Conference 2019, {DAC} 2019, Las Vegas, NV, USA, June
  02-06, 2019}.\hskip 1em plus 0.5em minus 0.4em\relax {ACM}, 2019, p. 160.

\bibitem{bfloat}
D.~D. Kalamkar \emph{et~al.}, ``A study of {BFLOAT16} for deep learning
  training,'' \emph{CoRR}, vol. abs/1905.12322, 2019.

\bibitem{dlfloat}
A.~Agrawal \emph{et~al.}, ``Dlfloat: A 16-b floating point format designed for
  deep learning training and inference,'' in \emph{2019 IEEE 26th Symposium on
  Computer Arithmetic (ARITH)}, 2019, pp. 92--95.

\bibitem{ternary}
S.~Zhu \emph{et~al.}, ``Tab: Unified and optimized ternary, binary, and
  mixed-precision neural network inference on the edge,'' \emph{ACM Trans.
  Embed. Comput. Syst.}, vol.~21, no.~5, oct 2022.

\bibitem{choi2018pact}
J.~Choi \emph{et~al.}, ``{PACT}: Parameterized clipping activation for
  quantized neural networks,'' 2018.

\bibitem{tcasZXSP19}
G.~Zervakis \emph{et~al.}, ``Multi-level approximate accelerator synthesis
  under voltage island constraints,'' \emph{{IEEE} Trans. Circuits Syst. {II}
  Express Briefs}, vol. 66-II, no.~4, pp. 607--611, 2019.

\bibitem{survey_ai_accl}
A.~Reuther \emph{et~al.}, ``Ai accelerator survey and trends,'' in \emph{2021
  IEEE High Performance Extreme Computing Conference (HPEC)}, 2021, pp. 1--9.

\bibitem{lascasLMXPS22}
V.~Leon \emph{et~al.}, ``Max-dnn: Multi-level arithmetic approximation for
  energy-efficient {DNN} hardware accelerators,'' in \emph{13th {IEEE} Latin
  America Symposium on Circuits and System, {LASCAS} 2022, Puerto Varas, Chile,
  March 1-4, 2022}.\hskip 1em plus 0.5em minus 0.4em\relax {IEEE}, 2022, pp.
  1--4.

\bibitem{8942068}
V.~Mrazek \emph{et~al.}, ``Alwann: Automatic layer-wise approximation of deep
  neural network accelerators without retraining,'' in \emph{2019 IEEE/ACM
  International Conference on Computer-Aided Design (ICCAD)}, 2019, pp. 1--8.

\bibitem{sparsity}
I.~Hubara \emph{et~al.}, ``Accelerated sparse neural training: A provable and
  efficient method to find n:m transposable masks,'' in \emph{Advances in
  Neural Information Processing Systems}, M.~Ranzato \emph{et~al.}, Eds.,
  vol.~34.\hskip 1em plus 0.5em minus 0.4em\relax Curran Associates, Inc.,
  2021, pp. 21\,099--21\,111.

\bibitem{SchumanPPBDRP17}
C.~D. Schuman \emph{et~al.}, ``A survey of neuromorphic computing and neural
  networks in hardware,'' \emph{CoRR}, vol. abs/1705.06963, 2017.

\bibitem{abs-2005-01467}
M.~Bouvier \emph{et~al.}, ``Spiking neural networks hardware implementations
  and challenges: a survey,'' \emph{CoRR}, vol. abs/2005.01467, 2020.

\end{thebibliography}


\begin{thebibliography}{10}
\providecommand{\url}[1]{#1}
\csname url@samestyle\endcsname
\providecommand{\newblock}{\relax}
\providecommand{\bibinfo}[2]{#2}
\providecommand{\BIBentrySTDinterwordspacing}{\spaceskip=0pt\relax}
\providecommand{\BIBentryALTinterwordstretchfactor}{4}
\providecommand{\BIBentryALTinterwordspacing}{\spaceskip=\fontdimen2\font plus
\BIBentryALTinterwordstretchfactor\fontdimen3\font minus
  \fontdimen4\font\relax}
\providecommand{\BIBforeignlanguage}[2]{{%
\expandafter\ifx\csname l@#1\endcsname\relax
\typeout{** WARNING: IEEEtran.bst: No hyphenation pattern has been}%
\typeout{** loaded for the language `#1'. Using the pattern for}%
\typeout{** the default language instead.}%
\else
\language=\csname l@#1\endcsname
\fi
#2}}
\providecommand{\BIBdecl}{\relax}
\BIBdecl

\bibitem{9138926}
S.~Narayanan~{\textit{et al.}}, ``Spinalflow: An architecture and dataflow
  tailored for spiking neural networks,'' in \emph{2020 ACM/IEEE 47th Annual
  International Symposium on Computer Architecture (ISCA)}, 2020, pp. 349--362.

\bibitem{fnins.2018.00774}
M.~Pfeiffer and T.~Pfeil, ``Deep learning with spiking neurons: Opportunities
  and challenges,'' \emph{Frontiers in Neuroscience}, vol.~12, 2018.

\bibitem{frenkel20180}
C.~Frenkel~{\textit{et al.}}, ``A 0.086-mm $\^{} 2 $12.7-pj/sop 64k-synapse
  256-neuron online-learning digital spiking neuromorphic processor in 28-nm
  cmos,'' \emph{IEEE transactions on biomedical circuits and systems}, vol.~13,
  no.~1, pp. 145--158, 2018.

\bibitem{kuang202164k}
Y.~Kuang~{\textit{et al.}}, ``A 64k-neuron 64m-1b-synapse 2.64 pj/sop
  neuromorphic chip with all memory on chip for spike-based models in 65nm
  cmos,'' \emph{IEEE Transactions on Circuits and Systems II: Express Briefs},
  vol.~68, no.~7, pp. 2655--2659, 2021.

\bibitem{wong20212}
M.~Wong~{\textit{et al.}}, ``A 2.1 pj/sop 40nm snn accelerator featuring
  on-chip transfer learning using delta stdp,'' in \emph{ESSDERC 2021-IEEE 51st
  European Solid-State Device Research Conference}, 2021, pp. 95--98.

\bibitem{chen20184096}
G.~K. Chen~{\textit{et al.}}, ``A 4096-neuron 1m-synapse 3.8-pj/sop spiking
  neural network with on-chip stdp learning and sparse weights in 10-nm finfet
  cmos,'' \emph{IEEE JSSC}, vol.~54, no.~4, pp. 992--1002, 2018.

\bibitem{zhang202128nm}
J.~Zhang~{\textit{et al.}}, ``A 28nm configurable asynchronous snn accelerator
  with energy-efficient learning,'' in \emph{2021 27th IEEE International
  Symposium on Asynchronous Circuits and Systems}.\hskip 1em plus 0.5em minus
  0.4em\relax IEEE, 2021, pp. 34--39.

\bibitem{teman2016power}
A.~Teman~{\textit{et al.}}, ``Power, area, and performance optimization of
  standard cell memory arrays through controlled placement,'' \emph{ACM
  Transactions on Design Automation of Electronic Systems (TODAES)}, vol.~21,
  no.~4, pp. 1--25, 2016.

\bibitem{akopyan2015truenorth}
F.~Akopyan~{\textit{et al.}}, ``Truenorth: Design and tool flow of a 65 mw 1
  million neuron programmable neurosynaptic chip,'' \emph{IEEE transactions on
  computer-aided design of integrated circuits and systems}, vol.~34, no.~10,
  pp. 1537--1557, 2015.

\bibitem{davies2018loihi}
M.~Davies~{\textit{et al.}}, ``Loihi: A neuromorphic manycore processor with
  on-chip learning,'' \emph{Ieee Micro}, vol.~38, no.~1, pp. 82--99, 2018.

\bibitem{painkras2012spinnaker}
E.~Painkras~{\textit{et al.}}, ``Spinnaker: A multi-core system-on-chip for
  massively-parallel neural net simulation,'' in \emph{Proceedings of the IEEE
  2012 Custom Integrated Circuits Conference}.\hskip 1em plus 0.5em minus
  0.4em\relax IEEE, 2012, pp. 1--4.

\bibitem{peeters2010click}
A.~Peeters~{\textit{et al.}}, ``Click elements: An implementation style for
  data-driven compilation,'' in \emph{2010 IEEE Symposium on Asynchronous
  Circuits and Systems}.\hskip 1em plus 0.5em minus 0.4em\relax IEEE, 2010, pp.
  3--14.

\bibitem{huang2020spiking}
X.~Huang~{\textit{et al.}}, ``Spiking neural network based low-power
  radioisotope identification using fpga,'' in \emph{2020 27th IEEE
  International Conference on Electronics, Circuits and Systems (ICECS)}.\hskip
  1em plus 0.5em minus 0.4em\relax IEEE, 2020, pp. 1--4.

\bibitem{mitchell2020small}
J.~P. Mitchell~{\textit{et al.}}, ``A small, low cost event-driven architecture
  for spiking neural networks on fpgas,'' in \emph{International Conference on
  Neuromorphic Systems 2020}, 2020, pp. 1--4.

\bibitem{irmak2021dynamic}
H.~Irmak~{\textit{et al.}}, ``A dynamic reconfigurable architecture for hybrid
  spiking and convolutional fpga-based neural network designs,'' \emph{Journal
  of Low Power Electronics and Applications}, vol.~11, no.~3, p.~32, 2021.

\bibitem{lee2021neuroengine}
H.~Lee~{\textit{et al.}}, ``Neuroengine: a hardware-based event-driven
  simulation system for advanced brain-inspired computing,'' in
  \emph{Proceedings of the 26th ACM International Conference on Architectural
  Support for Programming Languages and Operating Systems}, 2021, pp. 975--989.

\bibitem{stuijt2021mubrain}
J.~Stuijt~{\textit{et al.}}, ``$\mu$brain: An event-driven and fully
  synthesizable architecture for spiking neural networks,'' \emph{Frontiers in
  neuroscience}, vol.~15, p. 538, 2021.

\bibitem{wang2020always}
D.~Wang~{\textit{et al.}}, ``Always-on, sub-300-nw, event-driven spiking neural
  network based on spike-driven clock-generation and clock-and power-gating for
  an ultra-low-power intelligent device,'' in \emph{2020 IEEE Asian Solid-State
  Circuits Conference (A-SSCC)}.\hskip 1em plus 0.5em minus 0.4em\relax IEEE,
  2020, pp. 1--4.

\bibitem{8325231}
C.~Frenkel, J.-D. Legat, and D.~Bol, ``A compact phenomenological digital
  neuron implementing the 20 izhikevich behaviors,'' in \emph{2017 IEEE
  Biomedical Circuits and Systems Conference (BioCAS)}, 2017, pp. 1--4.

\bibitem{842110}
K.~Boahen, ``Point-to-point connectivity between neuromorphic chips using
  address events,'' \emph{IEEE Transactions on Circuits and Systems II: Analog
  and Digital Signal Processing}, vol.~47, no.~5, pp. 416--434, 2000.

\bibitem{brader2007learning}
J.~M. Brader~{\textit{et al.}}, ``Learning real-world stimuli in a neural
  network with spike-driven synaptic dynamics,'' \emph{Neural computation},
  vol.~19, no.~11, pp. 2881--2912, 2007.

\end{thebibliography}

\end{document}